\documentclass[%
 reprint,
preprintnumbers,
nofootinbib,
 amsmath,amssymb,
 prl
floatfix,
]{revtex4-1}
\usepackage{CJK}
\usepackage{graphicx}
\usepackage{dcolumn}
\usepackage{bm}
\usepackage[colorlinks=true,pdfstartview=FitV,breaklinks=true]{hyperref}
\usepackage[dvipsnames,table]{xcolor}
\hypersetup{urlcolor=BlueViolet,
	    citecolor=Plum,
	    linkcolor=PineGreen}

\usepackage{lipsum}	    
\usepackage{titlesec}
\titlespacing*{\section}{0pt}{0.9\baselineskip}{0.8\baselineskip}

\usepackage{times}

\begin{document}

\setlength{\abovedisplayskip}{3pt}
\setlength{\belowdisplayskip}{3pt}

\preprint{CPPC-2021-03}

\title{Coloured gravitational instantons, the strong CP problem and the companion axion solution. }
\author{Zhe Chen}
 \email{zche8090@uni.sydney.edu.au}
\author{Archil Kobakhidze}
 \email{archil.kobakhidze@sydney.edu.au}
\affiliation{Sydney Consortium for Particle Physics and Cosmology, \\
 School of Physics, The University of Sydney, NSW 2006, Australia 
}

\begin{abstract}
Quantum gravity introduces a new source of the combined parity (CP) violation in gauge theories. We argue that this new CP violation 
gets bundled with the strong CP violation through the coloured gravitational instantons. Consequently, the standard axion solution to the strong CP problem is compromised. Further, we argue that the ultimate solution to the strong CP problem must involve at least one additional axion particle.

\end{abstract}

\maketitle


\section{The strong CP problem}

The Yang-Mills gauge theories exhibit a multiplicity of topologically distinct vacua \cite{tHooft:1976rip, tHooft:1976snw}. Furthermore, quantum transitions between vacua have non-zero probability due to instantons \cite{Belavin:1975fg} and, thus, the "true" vacuum, known as the $\theta$-vacuum, is given by a superposition of those vacua. When applied to the gauge theory of strong interactions, the Quantum Chromodynamics (QCD), this intricate vacuum structure leads to the violation of the combined charge (C) and parity (P) invariance, the CP invariance. The CP violation in QCD, however, is at odds with the experimental limits on the neutron electric dipole moment, unless the relevant CP parameter, $\theta_{\mathrm{QCD}}$, is fine-tuned to a tiny value\footnote{Here $\theta_{\mathrm{QCD}}$ denotes the CP-violating parameter that includes a contribution from the electroweak sector.} \cite{Crewther:1979pi} (for a review, see, e.g., \cite{Peccei:2006as}), 
\begin{equation}
\theta_{\mathrm{QCD}}\lesssim 10^{-10}~.
\label{theta}
\end{equation}
 
The Peccei-Quinn mechanism \cite{Peccei:1977hh, Peccei:1977ur} is arguably the most elegant solution to the strong CP problem. It postulates gobal chiral symmetry $U(1)_{\mathrm{PQ}}$, the Peccei-Quinn symmetry, which exhibits a mixed $U(1)_{\mathrm{PQ}}$-QCD quantum anomaly, and the resulting violation of $U(1)_{\mathrm{PQ}}$ charge conservation is due to the same QCD instantons. In addition, the Peccei-Quinn  symmetry is broken spontaneously, producing a light pseudo-Goldstone boson, the axion \cite{Weinberg:1977ma, Wilczek:1977pj, Kim:1979if, Shifman:1979if, Zhitnitsky:1980tq, Dine:1981rt}. In the absence of any other source of $U(1)_{\mathrm{PQ}}$ symmetry breaking, the axion automatically develops the vacuum expectation value that cancels out $\theta_{\mathrm{QCD}}$, so that the vacuum state becomes CP-invariant. As a bonus to the solution of the strong CP problem, the axion represents a viable candidate for dark matter.   

It has been known for quite some time that gravitational instantons\footnote{For a comprehensive review of gravitational instantons,  see  Ref. \cite{Eguchi:1980jx}.} may support new CP-violating topological terms in the standard model \cite{Deser:1980kc, Holman:1992ah, Arunasalam:2018eaz}\footnote{Destabilising role of gravitational instantons is also discussed in Ref. \cite{Dvali:2005an} within the dual description of the strong CP problem}. For example, electrically charged Eguchi-Hanson (EH) gravitational instantons  \cite{Eguchi:1978xp, Eguchi:1978gw} induce the respective electromagnetic and gravitational $\theta$-terms in the effective Lagrangian of the standard model \cite{Holman:1992ah, Arunasalam:2018eaz}. The impact of this new CP-violation on the standard axion solution to the strong CP problem has been examined in Ref. \cite{Holman:1992ah} and found that the electromagnetic $\theta_{\mathrm{EM}}$ contribution to the effective axion potential is small enough (at least within the standard model) to affect the original solution\footnote{It has been also pointed out, that quantum gravity effects may explicitly break the global Peccei-Quinn symmetry and thus potentially jeopardise the axion solution to the strong CP problem \cite{Kamionkowski:1992mf}. Such breaking is typically accounted for within the effective field theory by high-dimensional operators with  \emph{a priory} unknown strengths. Some may argue that the operators are actually sufficiently suppressed to alter the Peccei-Quinn solution \cite{Kallosh:1995hi, Dvali:2013cpa}, or there could be additional theoretical input to avoid the "axion quality" problem (for a recent discussion see, e.g. \cite{Cox:2019rro}
}. 

However, there are known non-Abelian counterparts of the electromagnetic gravitational instantons \cite{BoutalebJoutei:1979rw, Chakrabarti:1987kz}, in particular, the QCD coloured Eguchi-Hanson (CEH) instantons. In this paper, we describe these instantons  and compute their contribution to the effective axion potential. We find that CEH instantons bundle the new gravity-induced source of CP-violation with the CP-violation in the sector of strong interactions in such a way that compromises the original axion solution to the strong CP problem.  Further, we propose an \emph{\`{a} la} Peccei-Quinn remedy to this re-emerged strong CP problem by introducing spontaneously broken anomalous  $U(1)'_{\mathrm{PQ}}$ in addition to the standard $U(1)_{\mathrm{PQ}}$. Consequently, in our scenario, the Peccei-Quinn axion is necessarily accompanied by a companion axion. The phenomenology of this companion-axion solution to the strong CP problem are discussed in \cite{Chen:2021hfq, Chen:2021wcf}.    
             
The paper is organised as follows. In the next section, we describe  asymptotically locally flat CEH instantons. In section \ref{axion}, we compute the CEH contribution to the effective axion potential and demonstrate that the standard Peccei-Quinn mechanism fails to solve the strong CP problem. Following this, in section \ref{companion}, we propose a remedy to this problem by introducing a companion axion in addition to the standard Peccei-Quinn axion. The last section \ref{conclusion} we reserve for conclusions. 

\section{Coloured gravitational instantons}
\label{instanton}

In this section, we recall the salient features of the combined Yang-Mills - gravitational instantons. First, we point out that since we are interested in processes described by
(asymptotically) flat spacetime S-matrix, we restrict ourselves to asymptotically Euclidean (AE) or asymptotically locally Euclidean (ALE) manifolds. The AE and ALE vacuum manifolds are known to be Ricci flat, $R = 0$, and have non-negative Einstein-Hilbert action, $S_{\mathrm{EH}}\geq 0$, according to the positive action theorem \cite{Schon:1979uj, Witten:1981mf}. Furthermore, while for AE manifolds $S_{\mathrm{EH}} = 0$ implies that they are Riemann-flat (no gravity), ALE manifolds with $S_{EH}=0$ support non-trivial (anti)self-dual Riemann curvature. Hence, it is reasonable to think that ALE vacuum manifolds support topologically non-trivial vacuum structure of a combined Yang-Mills-gravity theory in addition to the Yang-Mills instanton vacuum structure in flat spacetime.

The combined Yang-Mills - gravitational instantons  \cite{BoutalebJoutei:1979rw, Chakrabarti:1987kz} can be defined on multi-centred ALE spacetimes \cite{Gibbons:1978tef} and their topological properties are dramatically different from their flat spacetime counterparts. This is largely related to the fact that the infinity of ALE spacetime is topologically equivalent to $S^3/Z_N$ ($N=1$ is the usual flat spacetime). Hence, $SU(2)$ instantons in ALE spacetimes describe transitions between topologically distinct vacua that are classified homotopy of the mappings $S^3\to S^3$ modulo $Z_N$. In what follows we describe the simplest instanton on the EH ALE space ($N=2$), where the $SU(2)$ Yang-Mills configuration is defined through the EH spin-connection and hence its topological index is entirely determined by the topology of the base space. It turns out that this so-called "standard embedding" is the most relevant instanton for the strong CP problem and its axion solution. 


The metric for the pure gravitational EH instanton is given by:
\begin{eqnarray}
ds^2&=&\frac{1}{1-\frac{a^4}{r^4}}dr^2 + \frac{r^2}{4}\left[d\theta^2 +\sin^2\theta d\phi^2\right. \nonumber \\
&+&\left. \left(1-\frac{a^4}{r^4}\right)\left(d\psi+\cos\theta d\phi\right)^2\right]~,
\label{metric1}
\end{eqnarray}
where $r$ is the 4D radial coordinate and $(\theta, \phi, \psi)$ are angular coordinates; $a$ is an arbitrary integration constant interpreted as the instanton size. This instanton is known to correspond to the two-centre instanton of Ref. \cite{Gibbons:1978tef} (see the explicit coordinate transformations that relate two metrics in Ref. \cite{Prasad:1979kg})  

The metric (\ref{metric1}) is evidently singular at $r=a$. However, this  singularity can be removed by performing a $\mathbb{Z}_2$ identification of the coordinates. This can be seen as follows.
 Letting $u=r\sqrt{1-a^4/r^4}$, it can be shown that near $r=a$ (or $u=0$), the metric can be rewritten in terms of the Euler angles on $S^3$ as \cite{Eguchi:1978gw}: 
 \begin{align}
ds^2\simeq \frac{1}{4}\mathop{du^2}+\frac{1}{4} u^2 (\mathop{d\psi}+\cos\theta \mathop{d\phi})^2+\frac{a^2}{4}(\mathop{d\theta^2+\sin^2\theta\mathop{d\phi^2}})
\label{metric2} 
 \end{align}
Here, it is evident that at fixed $\theta$ and $\phi$, the metric becomes the usual metric of a plane with radial coordinate $u$ and angular coordinate $\psi$. Therefore, to remove the apparent singularity at $u=0$, one must restrict the domain of $\psi$ to $[0,2\pi)$. With this modification, it is clear that the topology of this space near the horizon, $r=a$ is that of $S^2\times \mathbb{R}^2$ where the sphere is parametrised by $(\theta,\phi)$ and the plane by $(u, \psi)$.  As $r\to\infty$, the metric asymptotically approaches that of flat spacetime but with the restriction in the domain of $\psi$, we see that the boundary at infinity is in fact $S^3/\mathbb{Z}_2=\mathbb{R}\mathrm{P}^3$.

The EH spacetime  (\ref{metric1}) is Ricci flat, i.e., $R=0$. One can easily verify that the Gibbons-Hawking boundary term is also vanishing and hence the total action is zero \cite{Eguchi:1978xp, Eguchi:1978gw}. This may create a false impression that the EH instanton could mediate unsuppressed anomalous processes with chiral charge violation. However, there are no normalisable fermion zero-modes in the background of the pure gravitational EH instanton, and hence such instantons have no relation to the chiral anomaly. 

Interestingly, the metric (\ref{metric1}) supports a self-dual  $SU(2)$ gauge field configurations, which are relevant to QCD once embedded into $SU(3)$. Among them, the focus of our interest 
here is the so-called "standard embedding" configuration, where $SU(2)$-valued gauge potential $A_{\mu}^a$ ($a=1,2,3$ are $SU(2)$ gauge indices) is expressed through the EH spin-connection $\omega_{\mu}^{AB}$ ($A,B=0,1,2,3$ are tangent space indices) as:  
\begin{align}
&A^a_{\mu}=\frac{1}{2}\eta_{AB}^{a}\omega_{\mu}^{AB}~, \nonumber \\
&\omega^{01}_{\theta}=\omega^{23}_{\theta}=\omega^{02}_{\phi}=\omega^{31}_{\phi}=\frac{1}{2}\sqrt{1-\frac{a^4}{r^4}}~, \nonumber \\
&\omega^{03}_{\psi}=\omega^{12}_{\psi}=\frac{1}{2}\left(1+\frac{a^4}{r^4}\right)~, 
\label{gauge}
\end{align}
where $\eta^a_{AB}$ is the self-dual 'tHooft symbol: $\eta^a_{A0}=\delta_{aA}$ and $\eta^a_{AB}=\epsilon_{aAB}$ for $A,B=1,2,3$. The above EH spin-connection is self-dual with respect to tangent space indices: $\omega_{\mu}^{AB}=-\frac{1}{2}\epsilon_{ABCD}\omega_{\mu}^{CD}$. Correspondingly, the Rieman tensor and the $SU(2)$ field strength are self-dual as well: $R_{\mu\nu\rho\sigma}=\tilde R_{\mu\nu\rho\sigma}=\frac{\sqrt{g}}{2}\epsilon_{\mu\nu\alpha\beta}g^{\alpha\gamma}g^{\beta\delta}R_{\gamma\delta\rho\sigma}$ and $F^a_{\mu\nu}=\tilde F^a_{\mu\nu}=\frac{\sqrt{g}}{2}\epsilon_{\mu\nu\alpha\beta}g^{\alpha\gamma}g^{\beta\delta}F^a_{\gamma\delta}$, respectively ($\epsilon_{\mu\nu\rho\sigma}$ denotes the  Levi-Chivita tensor with $\epsilon_{0123}=1$). Since the pure gravitational part of the action is zero for EH metric, the action for CEH instanton (\ref{gauge}) is defined entirely by the Yang-Mills part of the action: 
\begin{align}
S_{\rm CEH}&=\frac{1}{4g^2}\int d^4x \sqrt{g}F^{a\mu\nu}F^a_{\mu\nu}=\frac{4\pi^2}{g^2}\times 3~. 
\label{action}
\end{align}
Note the $4\pi^2$ factor in Eq. (\ref{action}) vs the standard $8\pi^2$ which appears in the action for pure Yang-Mills instantons. This is due $Z_2$ identification of antipodal points in the EH space, which becomes half of the (asymptotic) Euclidean space. We also note that, unlike the non-gravitational instantons, the final result in (\ref{action}) originates from the two boundary contributions evaluated at $r=a$ and  at $r=\infty$. 
 
The factor 3 in Eq. (\ref{action}) can be associated with the gravitational Pontryagin index which, due to the relations in (\ref{gauge}), is twice the $SU(2)$ gauge Pontryagin index:
\begin{eqnarray}
p&=&\frac{1}{16\pi^2}\int d^4x \sqrt{g}R_{\mu\nu\rho\sigma}\tilde R^{\mu\nu\rho\sigma} \nonumber \\
&=&\frac{2}{32\pi^2}\int d^4x \sqrt{g} F_{\mu\nu}^a\tilde F^{a\mu\nu}=3~.
\label{pontryagin}
\end{eqnarray}
The above Pontryagin index in turn is related to the topological Hirzebruch signature of EH spacetime, $\tau_{EH}=p/3 =1$. 

The vaccum-to-vaccum transitions mediated by CEH and BPST \cite{Belavin:1975fg} instantons are accounted by adding to the action $S[\Phi]$ two topological terms:
\begin{eqnarray}
S_{eff}[\Phi]&=&S[\Phi]+\frac{\theta_{QCD}}{32\pi^2}\int d^4x \sqrt{g} F_{\mu\nu}^a\tilde F^{a\mu\nu} \nonumber \\
&+&\frac{\theta_{EH}}{48\pi^2}\int d^4x \sqrt{g}R_{\mu\nu\rho\sigma}\tilde R^{\mu\nu\rho\sigma} 
\label{topterms}
\end{eqnarray}
The normalisation of the topological terms is fixed such that the BPST-instanton vacua are parametrised by $\theta_{QCD}$ and CEH-instanton vacua by $\theta_{CEH}=3\theta_{QCD}/2+\theta_{EH}$. 
   
In the presence of fermions, the transition amplitudes get modified. This is captured by another important topological characteristic of the CEH instantons the index of the Dirac operator. Following the generalised index theorem \cite{Atiyah:1975jf}, this index has been explicitly computed in \cite{Bianchi:1996zj} for a generic representation $R$ of $SU(2)$ of dimension $d_R$. Adopting these calculations for the "standard embedding" EH instantons, we obtain:
 \begin{eqnarray}
\nu^{(R)}_{1/2}=\left \lbrace 
\begin{tabular}{ll}
$\frac{1}{4}d_R(d_R^2-2)$~, & for $d_R=2,4,...$ \\ 
\\
$\frac{1}{4}d_R(d_R^2-1)$~,  & for $d_R=1,3,...$ \\ 
\end{tabular} 
\right.
\label{index}
\end{eqnarray}
We observe that for gauge-singlet fermions $d_R=1$ and $\nu^{(R)}_{1/2}=0$, hence pure gravitational EH instantons do not mediate chiral-symmetry breaking transitions; on the other hand, for the fermions (relevant for the Standard Model quarks)  in the fundamental representations $d_R=2$ the spin-1/2 fermion index is defined through the Hirzebruch signature, $\nu^{(R)}_{1/2}=\tau_{EH}=1$. This is the same as the index in the background of flat spacetime BPST instanton, although the Pontryagin indices of CEH and BPST instantons differ. Consequently, the number of fermionic zero-modes is the same in the backgrounds of CEH and BPST instantons and, hence, these instantons generate in the effective Lagrangian the chiral symmetry breaking terms of the form $\propto \mathrm{e}^{-\frac{12\pi^2}{g^2}}\mathrm{det}\left(\psi_L\bar \psi_R \right)$  and $\propto \mathrm{e}^{-\frac{8\pi^2}{g^2}}\mathrm{det}\left(\psi_L\bar \psi_R \right)$, respectively.   

The calculation of transition amplitudes requires the knowledge of the explicit form of CEH-instanton zero modes. From topological considerations, there are 12 bosonic zero modes \cite{Bianchi:1996zj} for $SU(2)$ CEH instanton, but they are not known explicitly. We will present more detailed analysis of zero-mode solutions elsewhere \cite{forthcoming2}, while here we estimate transition amplitudes based on the following educated guess. Since both  amplitudes generated by the ordinary QCD and CEH instantons have the same dimension, the Jacobian of collective coordinates of the CEH instanton must contain the same dimensional factor factor $\rho^{-5}d\rho$. The norm of each of 12 zero modes of $SU(2)$\footnote{We expect further 4 gauge zero modes from $SU(2)$ embedding into $SU(3)$ \cite{Bernard:1979qt, Vainshtein:1981wh}.} contributes a factor $(8\pi^2/g^2)^{1/2}$.  Finally, the renormalisation will introduce a factor $(\rho\Lambda)^{3b/2}$, where $b$ is the one-loop beta function coefficient and $\Lambda$ is the regularisation scale. Other numerical factors are expected to be the same order of magnitude as in the standard QCD calculations.  With this knowledge, we are ready now to compute the instanton-induced axion potential. 
  
\section{The axion potential}
\label{axion}

In the context of the standard model, the results from the previous section imply that we have two independent $\theta$ parameters, $\theta_{\mathrm{QCD}}$ and 
$\theta_{\mathrm{ CEH}}$ (\ref{topterms}), that parametrise topologically non-trivial vacuum structure\footnote{We ignore here possible electromagnetic and weak $\theta$ parameters, since their contributions are negligible for the problem we are interested in.}. It is important to note that even if one tunes $\theta_{QCD}$ to be zero, the relevant parameter that must satisfy the experimental constraint (\ref{theta}) becomes $\theta_{CEH}=\theta_{EH}$. Therefore,  any viable axion solution must ensure the cancellation of both CP-violating contributions to the desired accuracy (see also Ref. \cite{Dvali:2005an} for a related discussion). 

Let us consider a generic invisible axion model with the anomalous $U(1)_{\mathrm{PQ}}$ Peccei-Quinn symmetry being spontaneously broken at a high energy scale $f_a$. The leading contribution to the axion potential comes from the low-energy theory large-scale QCD and CEH instantons, where the relevant fermion degrees of freedom are $N_f=3$ light quarks $q_i~(i=u,d,s)$. The $|\nu_{\mathrm{QCD}}|=1$ pure QCD and $|\nu_{\mathrm{CEH}}|=1$ CEH instantons of sizes in the range $[\rho, \rho+d\rho]$ induce 6-fermion effective operators ('t Hooft vertices): 
\begin{align}
\Delta \mathcal{L}&=-d\left(\frac{2\pi}{\alpha_s(\Lambda)}\right)^6\mathrm{exp}\left[-\frac{2\pi}{\alpha_{s}(\Lambda)}+i\xi \frac{a}{f_a}+i\theta_{\mathrm{QCD}} \right] \nonumber  \\ 
&\times \frac{d\rho}{\rho^5}(\Lambda\rho)^{b}\mathrm{det}\left(q_{iL}\bar q_{jR}\right)\nonumber \\
&-\bar d\left(\frac{2\pi}{\alpha_s(\Lambda)}\right)^{8}\mathrm{exp}\left[-\frac{3\pi}{\alpha_{s}(\Lambda)}+i\bar \xi \frac{a}{f_a}+i\theta_{\mathrm{CEH}} \right] \nonumber \\
&\times \frac{d\rho}{\rho^5}(\Lambda\rho)^{3b/2}\mathrm{det}\left(q_{iL}\bar q_{jR}\right) +\mathrm{h.c.}
\label{effL}
\end{align}
Here, $d\simeq\frac{\mathrm{e}^{-3.7+0.26N_f}}{\pi^2}\approx 0.017$ is the renormalisation factor computed in the minimal subtraction scheme (we expect $\bar d\simeq d$), $\alpha_s (\Lambda)$ is the QCD running coupling evaluated at scale $\Lambda$ and $b=11-\frac{2}{3}N_f$ is the one-loop $\beta$-function that governs evaluation of $\alpha_s$ at one-loop: $\alpha^{-1}_{s}(\rho)=\alpha^{-1}_{s}(\Lambda)-\frac{b}{2\pi}\mathrm{ln}\left(\Lambda\rho\right)$; $\xi=\sum_q \left[2Q_{\rm PQ}C_2(R)\right]$ and  $\bar \xi=\sum_q \left[ 3Q_{\rm PQ}C_2(R)+Q_{\rm PQ}d_R/4\right]$
is the mixed $U(1)_{\mathrm{PQ}}$-QCD and the combined $U(1)_{\mathrm{PQ}}$-QCD-gravity anomaly coefficients, respectively, that depend on the Peccei-Quinn charge assignment of coloured fermions in a specific model.  We note that, as per standard approximation, we have ignored in (\ref{effL}) contributions from instantons with higher topological charge (see, e.g. \cite{Rennecke:2020zgb}), as well as contributions that steams from the overlap of QCD and CEH instantons. All these contributions will be subdominant in the standard dilute instanton gas approximation, where the instanton size is much smaller that average inter-instanton distance.      
   
To calculate axion potential we tie up the fermion (quasi)zero-modes via their masses, i.e. $\left(q_{iL}\bar q_{iR}\right)\to m_i\rho$ in (\ref{effL}) \cite{Vainshtein:1981wh}, and integrate over instanton size $\rho$ [note, $m_i\rho<<1$ in the dilute instanton approximation]. We obtain:
\begin{align}
V(a)&=-2K_{\mathrm{QCD}}\cos\left(\xi \frac{a}{f_a}+\theta_{\mathrm{QCD}}\right)\nonumber \\&-2K_{\mathrm{CEH}}\cos\left(\bar\xi \frac{a}{f_a}+\theta_{\mathrm{CEH}}\right)~,
\label{potential}
\end{align}    
where, 
\begin{align}
&K_{\mathrm{QCD}}=\frac{d}{8}\left(\frac{2\pi}{\alpha_s(\Lambda)}\right)^6\mathrm{e}^{-\frac{2\pi}{\alpha_s(\Lambda)}}m_um_dm_s\Lambda \nonumber \\
&=-\frac{m_um_dm_s \langle 0\vert\bar q q\vert 0\rangle}{m_um_d+m_um_s+m_dm_s}\approx \frac{m_{\pi}^2f_{\pi}^2m_um_d}{(m_u+m_d)^2}~,
\label{coef1} \\
&\frac{K_{\mathrm{CEH}}}{K_{\mathrm{QCD}}}\approx \frac{16}{25}\left(\frac{2\pi}{\alpha_s(\Lambda)}\right)^2\mathrm{e}^{-\frac{\pi}{\alpha_s(\Lambda)}}\approx 0.06 - 0.4~.
\label{coef2}
\end{align}
In Eq. (\ref{coef1}) we have equated instanton calculation of $K_{\mathrm{QCD}}$ with the same coefficient computed in chiral perturbation theory. The numerical values in Eq. (\ref{coef2}) 
are obtained by using the value of the strong coupling constant calculated in the minimal substruction scheme, $\alpha_s(\Lambda = 1 GeV)=0.4 - 0.6$ \cite{Deur:2016tte}. We observe that CEH contribution originates from large-size instantons and thus is not exponentially suppressed, unlike the contribution from the electrically charged instantons estimated in Ref. \cite{Holman:1992ah}. This has important ramifications for CP violation in QCD. It is easy to see that the axion ground state described by the minimisation of the axion potential (\ref{potential}) is not CP-invariant, unless two independent sources of CP violation are fine tuned, $\xi\theta_{\mathrm{CEH}}=\bar \xi\theta_{\mathrm{QCD}}$.  Consequently, the neutron EDM bound (\ref{theta}) is not satisfied.

\section{Companion axion solution to the strong CP problem}
\label{companion}
We saw in the previous section that the standard Peccei-Quinn mechanism with a single axion fails to resolve the strong CP problem. Following the footsteps of Peccei and Quinn, it is not difficult to find a remedy to this problem, however. Let us assume that the theory posses a new spontaneously broken anomalous symmetry $U(1)'_{PQ}$ in addition to the standard $U(1)_{PQ}$\footnote{We note in passing that in Ref. \cite{Dvali:2013cpa}, it was suggested that neutrino condensation and the associated "$\eta$-meson" can resolve the potential problem with gravitational anomalies.}. in the low energy limit the theory then would contain a new axion field $a'$, the companion field of the standard Peccei-Quinn axion $a$. The two-axion potential then reads: 
\begin{align}
V(a, a')&=-2K_{\mathrm{QCD}}\cos\left(\xi \frac{a}{f_a}+
\xi' \frac{a'}{f_{a'}}
+\theta_{\mathrm{QCD}}\right)\nonumber \\&-2K_{\mathrm{CEH}}\cos\left(\bar\xi \frac{a}{f_a}+\bar \xi' \frac{a'}{f_{a'}}+\theta_{\mathrm{CEH}}\right)~,
\label{potential2}
\end{align}
where $\xi'$ and $\bar \xi'$ are the anomaly coefficient associated with $U(1)'_{PQ}$ and $f_{a'}$ is the scale of $U(1)'_{PQ}$ breaking. The quick inspection of the above potential shows that its minimum is achieved for CP-conserving vacuum state, that is, the vacuum expectation values of the axion fields cancel out both $\theta$-terms. 

Note that two axion states are mixed through the interactions in Eq. (\ref{potential2}) [the mixing with neutral $\pi$ and $\eta$ mesons are neglected here] and hence do not represent mass eigenstates of physical axions. We readily compute the masses of physical axions, 

\begin{align}
    m^2_{\pm}&=K_{QCD}\left(\frac{\xi^2}{f^2_a}+\frac{\xi'^2}{f^2_{a'}}\right)+K_{CEH}\left(\frac{\bar{\xi}^2}{f^2_a}+\frac{\bar{\xi'}^2}{f^2_{a'}}\right)\nonumber \\
& \pm  \left[\bigg( K_{QCD}\left(\frac{\xi^2}{f^2_a}-\frac{\xi'^2}{f^2_{a'}}\right)+K_{CEH}\left(\frac{\bar{\xi}^2}{f^2_a}-\frac{\bar{\xi'}^2}{f^2_{a'}}\right)\bigg)^2\right. \nonumber \\  
&+ 16\left. \left( K_{QCD} \frac{\xi \xi'}{f_af_{a'}}
+K_{CEH}\frac{\bar{\xi}\bar{\xi'}}{f_af_{a'}} \right )^2  \right ]^{1/2}~,
\end{align}
and the mixing angle between two states: 
\begin{align}
   \tan2\alpha=\frac{2K_{QCD}\frac{\xi \xi'}{f_af_{a'}}+2K_{CEH}\frac{\bar{\xi}\bar{\xi'}}{f_af_{a'}}}{K_{QCD} \left(
 \frac{\xi^2}{f^2_a}-\frac{{\xi'}^2}{f^2_{a'}} \right)+K_{CEH}\left( \frac{\bar{\xi}^2}{f^2_a}-\frac{\bar{\xi'}^2}{f^2_{a'}}\right)}~.
\end{align}

There is a wide range of options for masses and mixing, including hierarchical masses with small or large mixing of the axion states. The couplings to the standard model fields can be worked out straightforwardly. Some of the related phenomenology is discussed in Refs.  \cite{Chen:2021hfq, Chen:2021wcf}.

\section{Conclusion}
\label{conclusion}
In this paper, we have argued that the standard Peccei-Quinn solution to the strong CP problem is compromised due to the new source of CP violation in gravity-QCD sector. It is well-known quantum gravity effects via gravitational instantons support new CP-violating topological terms in the effective lagrangian of the standard model \cite{Deser:1980kc, Holman:1992ah, Arunasalam:2018eaz}.  Namely, coloured gravitational instantons considered in this paper together with the flat spacetime QCD instantons support both QCD as well as gravitational $\theta$-terms (see, eq. (\ref{topterms})). Furthermore, CEH "standard embedding" instantons mediate vacuum-to-vacuum transitions that violate chiral and Peccei-Quinn global symmetries explicitly due to the combined QCD and gravitational anomalies, and hence result in the dependence of physical observables both on QCD and gravitational $\theta$-terms via $\theta_{\rm CEH}$.  The standard single axion model is incapable of removing both the standard $\theta_{\rm QCD}$ and the additional $\theta_{\rm CEH}$ from the physical observables (e.g., from the neutron electric dipole moment), and hence the standard Peccei-Quinn mechanism is compromised. 

In this regard, it is worth to compare the contribution of CEH instantons to the axion potential with other known quantum gravity induced contributions \cite{Holman:1992ah, Giddings:1987cg}. In Ref. \cite{Holman:1992ah}, the contribution of electromagnetic Eguchi-Hanson instantons were estimated and it has found that the dominant contribution comes from small-scale instantons of size $\sim 1/f_a$. The key reasons for that are (i) the running electromagnetic coupling becomes larger at small scales, thus reducing the WKB-suppression factor for instanton induced amplitudes  as well as (ii) the fact that the fermion zero-modes must be tied up through the Higgs boson loop in order to have non-vanishing contribution. Contrary to that, the dominant contribution to the axion potential comes from large-scale coloured Eguchi-Hanson instantons (similar to the standard QCD instantons) where the fermion zero modes are tied up by their masses. Simple comparison of WKB-factors reveal that electromagnetic Eguchi-Hanson instanton contribution is negligible (see, eq. (10) in Ref. \cite{Holman:1992ah}), while coloured Eguchi-Hanson instanton contribution is comparable with the pure QCD instanton contribution (see, eq. (\ref{coef2})). 

Another known quantum gravity contribution to the axion potential is related to the gravitational wormhole solutions \cite{Giddings:1987cg}. Wormholes also break the Peccei-Quinn symmetry explicitly and hence introduce additional terms in the axion potential.  However, the strength of this breaking is significant only for small-size wormholes, $L\sim 1/\sqrt{M_P f_a}$, and is generally negligible unless unless $f_a$ is too large, $f_a\gtrsim 10^{16}$ GeV  \cite{Alonso:2017avz}. We must stress again, that the CEH instantons in contrast dominate at low energies, and their effect cannot be offset by constraining  high-energy parameters. 

With these conclusions, we then have argued that the genuine solution to the strong CP problem   \emph{\`{a} la} Peccei-Quinn must involve an additional companion axion field. Such two-axion models have rich phenomenological and cosmological implications that were partially studied in \cite{Chen:2021hfq, Chen:2021wcf}.

\begin{acknowledgments}

This work of AK was partially supported by the Australian Research Council through the Discovery Project grant  DP210101636 and  by  Shota  Rustaveli National Science Foundation of Georgia (SRNSFG) through the grant DI-18-335.
\end{acknowledgments}

\end{document}